\documentclass[10pt]{iopart} 
\usepackage{iopams} 
\usepackage{setstack} 
\usepackage{graphicx}

\newcommand{\eqref}[1]{(\ref{#1})}

\newcommand{\ket}[1]{\vert{#1}\rangle}
\newcommand{\bea}{\begin{eqnarray}}
\newcommand{\eea}{\end{eqnarray}}

\begin{document}

\title[A short iterative scheme within the wave operator formalism]{Global integration of the Schr\"odinger equation: \\
a short iterative scheme within the wave operator formalism using discrete Fourier transforms}

\author{Arnaud Leclerc\S, Georges Jolicard\dag}
\address{\S\ SRSMC, UMR CNRS 7565, Universit\'e de Lorraine, 1Bd Arago 57070 Metz, France}
\address{\dag\ Institut Utinam, UMR CNRS 6213, Universit\'{e} de Franche-Comt\'{e}, Observatoire de Besan{\c c}on,
41 bis avenue de l'Observatoire, BP1615, 25010 Besan{\c c}on Cedex, France}

\eads{\mailto{arnaud.leclerc@univ-lorraine.fr}, \mailto{georges.jolicard@utinam.cnrs.fr}}

\begin{abstract}

A global solution of the Schr\"odinger equation for explicitly time-dependent 
Hamiltonians is derived by integrating the non-linear differential equation 
associated with the time-dependent wave operator. 
A fast iterative solution method is proposed in which, however, numerous integrals over time have to be evaluated. 
This internal work is done using a numerical integrator based on Fast Fourier Transforms (FFT). 
The case of a transition between two potential wells of a model molecule driven by intense laser pulses is used as an illustrative example. 
This application reveals some interesting features of the integration technique. 
Each iteration provides a global approximate solution on grid points regularly distributed over the full time propagation interval. 
Inside the convergence radius, the complete integration is competitive with standard algorithms, especially when high accuracy is required.

\end{abstract} 




\maketitle

\section{ Introduction}

The numerical solution of the Schr\"odinger equation $i\hbar \partial \Psi/\partial t =H \Psi$ plays a key role
in the understanding of molecular dynamics processes. 
Molecular inelastic \cite{Sun1} and reactive collisions \cite{Neu1} 
have long been treated using quantal wave packet methods. 
In procedures such as the semiclassical treatment of multiphoton processes \cite{atabek2010}
explicitly time-dependent Hamiltonians have to be used to describe the molecule-field interaction. 
A popular and efficient wave-packet propagation algorithm is the Multi-Configuration Time-Dependent Hartree method,
which is able to handle a large variety of high-dimensional problems \cite{MCTDH}. 
Here, our aim is not to deal with many dimensions but to develop an efficient way of treating problems with rapidly 
oscillating time dependencies in the Hamiltonian operator or in the wavefunctions. 
Several algorithms which are used for time-resolved experiments with time-independent Hamiltonians can be adapted
to the time-dependent case and are described in \cite{Lefores1} and references therein. 
However for such methods
the length of the integration steps must be reduced to handle the high frequency terms. 
The propagation scheme
is then based on the decomposition of the evolution operator using small time increments. 
\begin{equation}
U(t,0)=\prod_{n=0}^{N-1}U\left((n+1)\Delta t,n\Delta t \right)
\label{evolution_op}
\end{equation}
where $ \Delta t=t/N$ and
\begin{equation}
U(t+\Delta t,t) \simeq \exp[-(i/\hbar)H(t+\Delta t /2) \; \Delta t]
\label{evolution_op_2}
\end{equation}
These step by step integration schemes will be referred to in later sections as "continuous methods". 
Second order differencing schemes (SOD) \cite{askar}, 
split operator methods \cite{feit}
and the short iterative Lanczos propagation \cite{park}
all give cumulative propagation errors which are proportional to
$(\Delta t)^3$ in \eqref{evolution_op} and also errors due to the approximate calculation of the action of $\exp[-(i/\hbar) H \Delta t]$
on the wave function. 
Other high-accuracy integrators which can be used for multi-dimensional systems include the generalization
of the sympletic partitioned Runge-Kutta method developed by Sanz-Serna and Portillo \cite{Sanz1} and the method proposed by
Kormann \cite{Kor1} in which the Hamiltonian $H$ is replaced by a suitable truncation of the Magnus series. 
Unfortunately, these continuous integrators suffer from limitations which could prevent their use in some cases. 
SOD cannot handle non-Hermitian Hamiltonians, 
while the split operator scheme requires that the kinetic operator does not mix coordinates and their
associated momenta.
The calculation of the Magnus development is only tractable if the couplings have
separate time and coordinate dependencies \cite{Kor1}. 

A recently proposed global treatment,
the Constrained Adiabatic Trajectory Method (CATM), \cite{Jol1,Lecl1,Lecl2,Lecl3} introduces the Floquet Hamiltonian
$H_F=H -i\hbar \partial/\partial t$ defined in an 
extended Hilbert space including the time coordinate. 
The dynamical Schr\"odinger equation is transformed into an equivalent partial eigenvalue problem 
in the extended space and the wave function is forced to conform to consistent boundary conditions over a short artificial time extension by using a time-dependent absorbing potential. 
In \cite{Lecl3}, the CATM was formulated 
 as a global integrator of the Schr\"odinger equation 
which does not belong to the category of methods described by \eqref{evolution_op} and \eqref{evolution_op_2}
and it was concluded that the CATM is well suited to the description of systems driven by Hamiltonians with explicit and complicated time variations.
The method does not suffer from cumulative errors (because of its global character)
and 
the only error sources are the non-completeness of the finite molecular and temporal basis sets used 
and the imperfection of the time-dependent absorbing potential. 
The dynamics of a quantum system can also be obtained in a global way (as in the CATM) by using the time-dependent wave operator concept (TDWO) \cite{Jol3}. 
In this formalism the dynamics is formally separated into a simple evolution within a given active subspace driven by an effective low-dimensional Hamiltonian plus a secondary evolution from the subspace to the complete space. 
The TDWO concept has also been generalized \cite{Vien1} to treat almost adiabatic quantum dynamics. 
This generalization is based on a time-dependent adiabatic deformation of the active space and
is useful for dynamics which do not escape too far from an adiabatic subspace. 
However the effectiveness of both TDWO and CATM hinges on the capacity to integrate some nonlinear differential equations and
this is the central problem treated in this work. 
A solution of this problem is essential to preserve the principal advantage of these formalisms, i.e.
their capacity to give a global solution over the whole interaction time. 
To reach this goal an iterative solution seems appropriate. 
Such a solution was proposed in ref. \cite{Jol2}, using a time-dependent version of the
Recursive Distorted Wave Approximation (RDWA) \cite{Jol3} together with non-linear transformations such as Pad\'e approximants
\cite{Kill1}. 
Nevertheless, this iterative solution is not fully satisfactory and often fails because of a small radius of convergence 
and a high sensitivity to the choice of the initial guessed solution. 

The present paper proposes a new iterative solution of the time-dependent wave operator equations. 
This is equivalent to solving the time-dependent Schr\"odinger equation {\emph{without}} using the approximation of \eqref{evolution_op} and \eqref{evolution_op_2} at any stage of the calculation. 
In section \ref{iterative_proc} the basic wave operator equations are presented together with the iterative
integration procedure. 
In practice we limit ourselves to the non-degenerate case (one-dimensional active subspace) which describes the dynamics issuing from a given quantum state. 
This is equivalent to solving the Schr\"odinger equation for the wavefunction. 
The iteration procedure involves calculating numerous integrals over time of matrix and vector elements. 
Section \ref{integrals} explores the possibility of using Fast Fourier Transforms (FFT) to compute these integrals.
In section \ref{results} we illustrate the algorithm by studying the transition of a vibrational wave packet between the two wells of a model potential energy surface under the influence of two laser fields.
Various numerical features are analysed in details. 
Section \ref{conclusion} gives a discussion and conclusion. 

\section{Iterative calculation of the time dependent wave operator \label{iterative_proc}}%

\subsection{Wave operator equations} 

Let $\mathcal{H}$ be the Hilbert space associated with a molecular system and let $S_o$ be a model  subspace of rank $m$.
It is assumed that the initial molecular state is included within $S_o$. 
The orthogonal projector corresponding to the model space is called $P_o$, with 
$P_o^2=P_o$, $P_o^{\dag}=P_o$, $tr(P_o)=m$. 
During the evolution, the model space is continuously transformed into successive active spaces $S(t)$ 
whose projectors $P(t)$ are solutions of the Schr\"odinger-von Neumann equation:
%
%
\begin{equation} 
i\hbar \frac{\partial P(t)}{\partial t}=[H(t),P(t)].
\label{schrod_proj}
\end{equation}
The wave operator associated with the two subspaces $S_o$ and $S(t)$ is defined as \cite{Jol3} 
%
%
\begin{equation}
\Omega (t)=P(t)(P_oP(t)P_o)^{-1}=U(t,0;H)(P_oU(t,0;H)P_o)^{-1}
\label{WOD}
\end{equation}
where $U$ represents the quantum evolution operator associated with the Hamiltonian $H(t)$, i.e.
$i\hbar  \frac{\partial U(t,0;H)}{\partial t} = H(t) U(t,0;H)$. 
In \eqref{WOD},
$(P_oPP_o)^{-1}$ is the inverse of $P$ within $S_o$. It exists only if the
Fubini-Study distance between $P_o$ and $P$ is small: $dist_{FS}(P_o,P) \leq \frac{\pi}{2}$ \cite{Vien1}, i.e.
if the dynamics does not escape too far from the initial subspace.
In this framework the time evolution can be written as 
%
%
\begin{eqnarray}
\left \lbrace \begin{array}{l}
U(t,0;H)P_o=\Omega (t) U(t,0;H_{eff}) \\
H_{eff}(t)=P_o H(t) \Omega (t)
\end{array} \right .
\label{TOTAL}
\end{eqnarray}
In equation \ref{TOTAL} the evolution issuing from the initial state is separated into two terms.
The first one, $U(t,0;H_{eff})$, describes 
the dynamics within the model space $S_o$. 
The second one can be written as $\Omega (t)=P_o +X(t)$ and it induces transitions from the model space to the full
space, the off diagonal part ($X=Q_o \Omega P_o$) inducing transitions to the complementary
space exclusively ($Q_o$ being the projector associated with the
complementary space). 
It is evident that the factorisation of 
$U(H_{eff})$, which includes all the fast variations within $S_o$, 
is an advantage for the dynamical integration process 
if the model space is well chosen.
It can be proven that the reduced wave operator $X(t)=Q_o \Omega P_o$ satisfies \cite{Jol3}
%
\begin{equation}
i\hbar \frac{\partial X(t)}{\partial t}=Q_o(1-X(t))H(t)(1+X(t))P_o.
\label{eq_diff}
\end{equation}
Solving this equation is the central point of the wave operator formalism. 
The solution gives 
$\Omega (t)$ as well as $H_{eff} (t)$ (cf equation \ref{TOTAL}),
from which a partial evolution operator can be obtained. 

A continuous integration of this non-linear differential equation has been proposed in section VI of the
review \cite{Jol3}. Unfortunately, many approximations were necessary to make
the calculation tractable and the final result was not satisfactory in the strong coupling regime.
To derive a new global solution, 
we can use the fact that $Q_o(1-X) \frac{\partial }{\partial t} (1+X)P_o = Q_o \frac{\partial }{\partial t} X P_o$ to rewrite \eqref{eq_diff} as
%
\begin{equation}
Q_o(1-X(t))H_F(t)(1+X(t))P_o=0
\label{eq7_hf}
\end{equation}
where $H_F(t)$ is the Floquet Hamiltonian,  
%
\begin{equation}
H_F(t)=H(t)-i\hbar \frac{\partial}{\partial t}.
\end{equation}
We now derive an iterative treatment, assuming that eq \eqref{eq7_hf} is not perfectly satisfied at a finite iteration order $n$, 
%
\begin{equation}
Q_o(1-X^{(n)}(t))H_F(t)(1+X^{(n)}(t))P_o=\Delta^{(n)}(t)\ne 0.
\label{INPE}
\end{equation}
We are looking for the increment $\delta X^{(n)}$ which exactly solves the problem:
%
\begin{equation}
Q_o(1-X^{(n)}(t)-\delta X^{(n)}(t))H_F(t)(1+X^{(n)}(t)+\delta X^{(n)}(t))=0.
\label{increment}
\end{equation}
Expanding \eqref{increment} gives four terms of different significance,
\begin{eqnarray}
\Delta^{(n)} - Q_o \delta X^{(n)}(t) H_F(t) \delta X^{(n)}(t) P_o &&  \nonumber \\
- Q_o \delta X^{(n)}(t) H_F(t) (1+X^{(n)}(t)) P_o &&\nonumber  \\
+Q_o(1-X^{(n)}(t)) H_F(t) \delta X^{(n)}(t) P_o &=& 0
\label{four_terms}
\end{eqnarray}
In \eqref{four_terms}, the first term is the error at the previous iteration defined in \eqref{INPE}. 
The second term is quadratic in the increment $(\delta X^{(n)}(t))$ and can be neglected;
this is an approximation commonly used in such iterative treatments. 
The third term can be rewritten as
\begin{eqnarray}
&& Q_o \delta X^{(n)} H_F (1+X^{(n)}) P_o \nonumber \\
&=& Q_o \delta X^{(n)} \left[ P_o H (P_o + Q_o X^{(n)} P_o) - P_o i \hbar \frac{\partial}{\partial t} (P_o + Q_o X^{(n)} P_o) \right] \nonumber \\
&=& Q_o \delta X^{(n)} P_o H (P_o + Q_o X^{(n)} P_o)   \nonumber \\
&=& Q_o \delta X^{(n)}(t) H_{eff}^{(n)}(t) P_o,
\end{eqnarray}
with
\begin{equation}
H^{(n)}_{eff}(t)=P_oH(t)(P_o+X^{(n)}(t)).
\label{Heff}
\end{equation}
The last term in \eqref{four_terms} needs more detailed attention: 
\begin{eqnarray}
&& Q_o(1-X^{(n)}(t)) H_F \delta X^{(n)}(t) P_o \nonumber \\
&=& Q_o H \; \delta X^{(n)} P_o - Q_o X^{(n)} H \; \delta X^{(n)}
i \hbar \frac{\partial}{\partial t} Q_o X^{(n)} P_o
\end{eqnarray}
We cannot handle differential equations in which $(H \delta X^{(n)})$ or $(X^{(n)} H)$  couple the unknown components 
of $\delta X^{(n)}$ to all the complementary space spanned by $Q_o$. 
We shall thus assume that the non-diagonal elements of $H$ in the complementary space, given by $Q_o(H-H_{diag})Q_o$, are 
small and make two last approximations, 
\begin{eqnarray}
&& Q_o H(t) Q_o   \approx  Q_o \left[ H(t) \right]_{diag} Q_o \nonumber \\
&& Q_o X^{(n)}(t) H(t) Q_o  \approx  Q_o \left[ X^{(n)}(t) H(t) \right]_{diag}Q_o 
\label{1app}
\end{eqnarray}
This simplification is important but does not have important consequences  
if the procedure still converges. 
The neglected terms are gradually taken into account during the iterative process and finally equation \eqref{eq7_hf} is satisfied.
In summary, 
the increment defined in \eqref{increment} can be computed by solving
\begin{equation}
i\hbar \frac{\partial}{\partial t}\delta X^{(n)}(t) = \Delta^{(n)}(t)-\delta X^{(n)}(t) H_{eff}^{(n)}(t)
+ \tilde{H}^{(n)}_{diag}(t) \; \delta X^{(n)}(t)
\label{1diff}
\end{equation} 
with $H_{eff}$ defined in \eqref{Heff} and
\begin{equation}
\tilde{H}^{(n)}_{diag}(t)=Q_o \left[ H(t)-X^{(n)}(t)H(t) \right]_{diag}Q_o
\label{SOLRIG1}
\end{equation}
In the following we limit ourselves to the case of a one-dimensional model subspace, with
$P_o=|i\rangle \langle i|$ and $Q_o=\sum_{k\neq i} \vert k \rangle \langle k \vert$, $|i\rangle$ being the initial 
state of the system. 
$X^{(n)}$ and $\delta X^{(n)}$ becomes vectors and $H_{eff}(t)$ is a pure scalar function. 
Equation \eqref{1diff} becomes an ordinary differential equation whose rigorous solution for the component $ \langle k \vert \delta X^{(n)} \ket{i} = \delta X^{(n)}_k$ can be expressed using exponential evolution operators, 
\bea
& \delta X^{(n)}_k (t) &=
\exp \left[ \frac{1}{i\hbar} \int_0^t \left( \tilde{H}^{(n)}_{diag,kk}(t') - H_{eff}^{(n)}(t') \right) dt'  \right]   
\label{SOLRIG} \\
& \times & 
\int_0^t \left\{ \Delta_k^{(n)} (t') \exp \left[ - \frac{1}{i \hbar} \int_0^{t'} \left( \tilde{H}^{(n)}_{diag,kk}(t'') - H_{eff}^{(n)}(t'') \right) dt'' \right] \right\} dt'.    \nonumber
\eea
This solution is consistent with the initial value of the wave operator 
\begin{equation}
\Omega(t=0) = P_o,
\end{equation}
hence $X^{(n)}(0)=0$. 
Equations \eqref{Heff}, \eqref{SOLRIG1} and \eqref{SOLRIG} are the central equations of this paper which are to be solved numerically.

\subsection{Periodic basis set and constraints on the boundaries \label{abs_pot}}

Similar equations to those given above were derived in \cite{Jol4} 
by expanding the evolution operator using successive interaction pictures. 
However 
they were not applied numerically because of the difficulty of calculating the many time integrals 
which are similar to those present in \eqref{SOLRIG}. 
Only adiabatic limits could be calculated, with very poor results. 
Here we show that it is possible to design an efficient implementation by observing 
that the global approach represented by equations \eqref{eq7_hf} and \eqref{SOLRIG} is consistent with the use
of a discrete Fourier basis set to span the whole propagation time interval, together with 
the associated periodic quadrature. These tools give effective ways to express
the matrices $H(t)X^{(n)}(t)$ and $\partial X^{(n)}(t)/\partial t$ present in
$\Delta^{(n)}(t)$ (equation \eqref{INPE}) and to calculate the integrals in \eqref{SOLRIG}, 
as will be explained in section \ref{integrals}. 
However the accurate use of Fast Fourier Transforms (FFT) is only possible if all the time-dependent functions are made perodic and smooth.

The Hamiltonian $H(t)$ of a molecule submitted to a laser pulse
is the sum of a time-independent part $H_o$ which represents the
unperturbed molecule and of a 
perturbation $V(t)$ with $V(0)=V(T_o)=0$, where $T_o$ is the
laser pulse duration. 
In equation~\eqref{INPE} 
this corresponds to $X^{(n)}(t=0)=0$ and thus $\Delta^{(n)}(t=0)=0$. 
However at the end of the pulse duration, $X^{(n)}(t=T_o)$ can differ considerably from its initial value. 
To impose the continuity and the initial conditions of $X^{(n)}(t)$ one should use the CATM procedure \cite{Jol1,Lecl1}. 
To this end, a time-dependent absorbing potential is introduced over an artificial time extension $[T_o,T]$. 
In the present case, this imaginary potential $V_{abs}(t)$ destroys all the components of 
the wave operator (and the wavefunction), except the one corresponding to the initial state $\langle i | \Omega(t)\rangle$,
%
%
\begin{equation}
V_{abs}(t)=-iV_{opt}(t)\sum_{k \ne i}|k\rangle\langle k|
\label{ABS}
\end{equation}
where $V_{opt}(t)$ is a real positive function localised on the interval $[T_o,T]$. 
This supplementary term produces the boundary value $ X^{(n)}(T) \rightarrow 0 $ and hence the periodicity of $\delta X^{(n)}$ and $\Delta^{(n)}$ (equation \eqref{INPE}):
%
\bea
\delta X^{(n)}(t=0) &=&\delta X^{(n)}(t=T)=0 , \nonumber \\
\Delta^{(n)}(t=0) &=&\Delta^{(n)}(t=T)=0 .
\eea

\subsection{The problem of nested integrals over time}

The $n^{th}$ order iteration requires solving the equations \eqref{Heff}, \eqref{SOLRIG1} and \eqref{SOLRIG}
and the passage to the $(n+1)^{th}$ order, via
the incrementation of $X$ using $X^{(n+1)}=X^{(n)}+\delta X^{(n)}$ 
and the construction of $\Delta^{(n+1)}$ (cf equation \eqref{INPE}). 
We want to use the same discretization at regularly distributed grid points to describe all the time dependencies. 
Suppose that the molecular basis set is composed of $N_v$ orthonormal states with $P_o=|i\rangle \langle i|$
and $Q_o=\sum_{k \ne i} |k\rangle \langle k |$ and that the discrete time-grid which spans the time-interval
$[0,T]$ is composed of $N$ discrete time values $t_j=jT/N, \; j=0,\ldots,N-1$.
The calculation of 
$\delta X^{(n)}(t)$ in \eqref{SOLRIG} calls for the calculation of 
$N_v-1$ integrals
$ \int_0^{t_j} \left( \tilde{H}^{(n)}_{diag,kk} - H_{eff}^{(n)} \right) dt' $,
where $k$ takes all the $N_v-1$ index values in the complementary space,
followed by $N_v-1$ integrals $\int_0^{t_j}  \Delta_j^{(n)}(t') e^{ - \frac{1}{i \hbar} \int_0^{t'} \left( \tilde{H}^{(n)}_{diag,kk} - H_{eff}^{(n)} \right) dt''}dt'.$
This is obviously a significant task, which must be undertaken at each interation order $(n)$. 
The calculation of each integral must be fast and very accurate. 
The selected procedure is explained in the next section with a preliminary discussion about integrating an arbitrary complex function of time.

\section{Discrete Fourier integration at the $n^{th}$ iteration order \label{integrals}}
In this section we explain how to compute efficiently a numerical definite integral
of a time-dependent function described on regularly distributed sampling points.  
The function is not necessarily continuous at the boundaries of the grid. 
This preliminary discussion is important in facilitating the numerical implementation of equation \eqref{SOLRIG}.
We require that the procedure should determine with the same accuracy, 
not only the integral at the final time value $t=T$ 
(which is easily obtained using the standard Fourier quadrature rule) 
but also its value for each of the $N$ intermediate discrete time values $t_j$.

\subsection{Continuous integrals and numerical FFT-integrators \label{theorie}} 
In the present subsection the Fourier transform $F(\nu)$ is defined as
%
%
\begin{equation}
F(\nu)=\int_{-\infty}^{+\infty} f(t) \exp (-i2\pi\nu t) dt.
\end{equation}                                             
%
%
Let $I(t)$ be the integral
\begin{eqnarray}
I(t)&=&\int_{-\infty}^{t}f(t') dt' \nonumber \\
& =& \int_{-\infty}^{+\infty} f(t') h(t-t') dt'
\end{eqnarray}
where $h(t)$ is the Heaviside step function, 
$$
h(t)=
\left \{
\begin{array}{l}
0 \;\; t<0 \\
1 \;\; t\ge 0.
\end{array} \right.$$ 
The Fourier transform of $h(t)$ is the distribution
%
%
\begin{equation}
H(\nu)=\frac{1}{2} \delta(\nu)-PV \left(\frac{ i }{2 \pi\nu} \right).
\label{fftheaviside}
\end{equation}
PV denotes the Cauchy principal value which becomes relevant when the above expression is integrated. 
The convolution theorem gives
%
%
\begin{eqnarray}
I(t) &=& f * h (t) \nonumber \\
	&= & \int_{-\infty}^{+\infty} F(\nu ) H(\nu) \exp( 2 i \pi \nu)  d\nu.
\end{eqnarray}
Using $H(\nu)$ from equation \eqref{fftheaviside} then gives 
%
%
\begin{equation}
I(t)  
=\frac{1}{2}F(0) - \frac{i}{2\pi} PV\int_{-\infty}^{+\infty}\frac{F(\nu)e^{2i\pi\nu t}}{\nu} d\nu.
\label{solution_continue}
\end{equation}
In molecule-field interaction problems, the field term is frequently an oscillating function of time. 
Let us consider for a moment the special  case of an oscillating function of the form 
$f(t) = g(t) \exp( i 2\pi \tilde{\nu} t)$. 
Its Fourier transform is the translated function
$$
F(\nu) = G(\nu - \tilde{\nu} ),
$$
where $G (\nu)$ is the transform of $g(t)$. 
Using this $F(\nu)$ in \eqref{solution_continue} and using the variable change $\nu' = \tilde{\nu} - \nu$, we obtain 
%
%
\begin{eqnarray}
I(t) &=& \int_{-\infty}^{t} g(t) \exp( i 2\pi \tilde{\nu} t)  dt = \frac{1}{2} G(-\tilde{\nu}) \nonumber \\
&-&\frac{i}{2\pi} e^{i2\pi \tilde{\nu} t} PV \int_{-\infty}^{+\infty}\frac{G(-\nu') e^{-2i\pi\nu' t}}{\tilde{\nu}-\nu'} d\nu'.
\label{solution_avec_exp}
\end{eqnarray}
When the frequency $\tilde{\nu}$ is very high, we can 
neglect the Fourier coefficient $G(-\tilde{\nu})$ and 
approximate the denominator $\tilde{\nu}-\nu' $ by $\tilde{\nu}$, giving the approximation 
%
%
\begin{equation}
\int_{-\infty}^t g(t') \exp( 2 i \pi \tilde{\nu} t')dt' \underset{\tilde{\nu} \to \infty} \thicksim     \exp[i( 2\pi \tilde{\nu} t-\pi/2)]\frac{g(t)}{2 \pi \tilde{\nu}}.
\label{ADAP}
\end{equation}

We can now go back to the general case of equation \eqref{solution_continue} to derive a numerical FFT-integrator. 
The detailed derivation is given in \ref{app_FFT}. 
This appendix explains how equation \eqref{solution_continue} can be implemented for periodic or non-periodic functions known as $N$ sample values distributed over the interval $[0,T]$. The integration algorithm needs only two FFT, with a computational cost scaling as $2 N \log N$ to obtain the $N$ values of $I(t_j)$, with $t_j= jT/N, j=0,1,\dots,N-1$. Here we recall only the main result, 
\begin{equation}
I(t_j)=FFT_j^{-1}(\mu.FFT(f)) + \frac{j}{N}\times \frac{T}{\sqrt N} FFT_{k=0}(f).
\label{eq_finale_1d_txt_princ}
\end{equation}
where $\mu.FFT (f)$ is a component by component vector multiplication,
with 
\begin{equation}
\mu_{\ell}=\left \{ \begin{array}{l}
\frac{i}{2\pi}\frac{T}{\ell} \;\;\;\; \ell=1, \ldots \frac{N}{2}-1  \; ,\\
\frac{i}{2\pi}\frac{T}{\ell-N} \;\;\;\; \ell=\frac{N}{2}, \ldots N-1. \\
\end{array} \right. \\
\label{bons_coefs_txt_princ}
\end{equation}
In addition, the $\ell=0$ component of $(\mu . FFT(f))$ in \eqref{eq_finale_1d_txt_princ} must be replaced by the following quantity:
\begin{equation}
a =  - \sum_{\ell=1}^{N-1} \mu_{\ell}.FFT_{\ell} (f). 
\label{coef_manquant_txt_princ }
\end{equation}
In the next subsection we perform some numerical tests using the algorithm derived in appendix \ref{app_FFT} with rapidly oscillating functions. 

\subsection{Numerical test of the FFT-based integration algorithm \label{numerique}}
\begin{figure}[htp]
\centering
\includegraphics[width=0.7\linewidth]{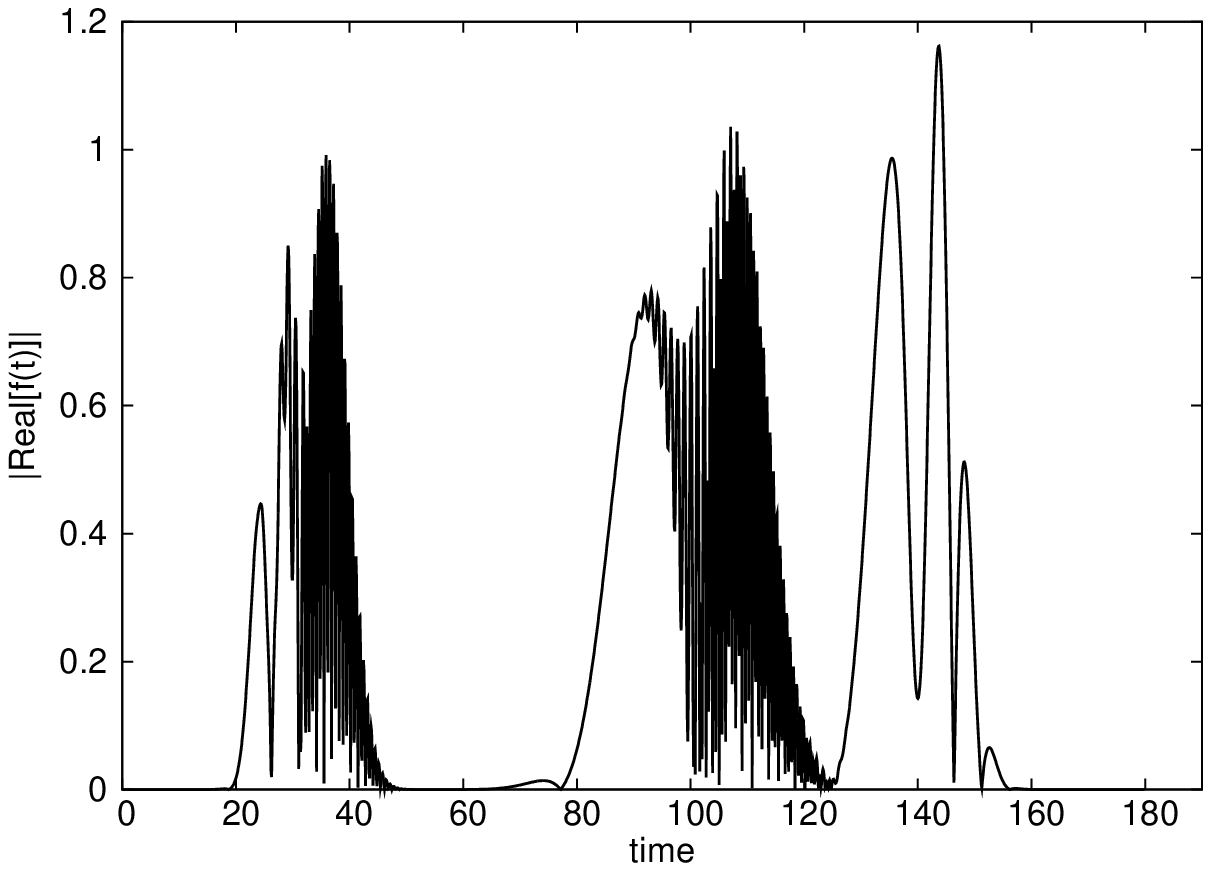}
\caption{$\vert \Re e f(t) \vert$, see \eqref{ex2} and table \ref{parameter1}.}
\label{FoncEx2}
\end{figure}

We study the arbitrary case of a function composed of six gaussians multiplied 
by complex exponential functions.
The gaussians have various widths and are distributed on the interval $[0,T=180]$ to give the function 
%
%
\begin{equation}
f(t)=\sum_{j=1}^6 \exp[-a_j (t-t_j)^2] \exp(i 2\pi x_j t/T).
\label{ex2}
\end{equation}
The various parameters appearing in \eqref{ex2} are given in table \ref{parameter1}.
\begin{table}[htp]
\centering
\begin{tabular}{llll}
\hline
\hline
Gaussian &\;\; $a_j$ & $t_j$ & $x_j$  \\
index $j$ & & &  \\
\hline
1 &6.790&27.000 &12.000 \\
2 &3.819&36.000 &135.600 \\
3 &1.018&90.000 &1.750 \\
4 &1.591&108.000 &154.700 \\
5 &2.118&135.000 &3.250 \\
6 &3.310&144.000 &18.150 \\
\hline
\end{tabular}
\caption{Parameters used in equation \eqref{ex2}.}
\label{parameter1}
\end{table}
The $a_j$ are chosen so that the modulus of the function $f(t)$ is equal to or smaller than $10^{-20}$
at the two boundaries (see figure \ref{FoncEx2}).
\begin{figure}[htp]
\centering
\includegraphics[width=0.7\linewidth]{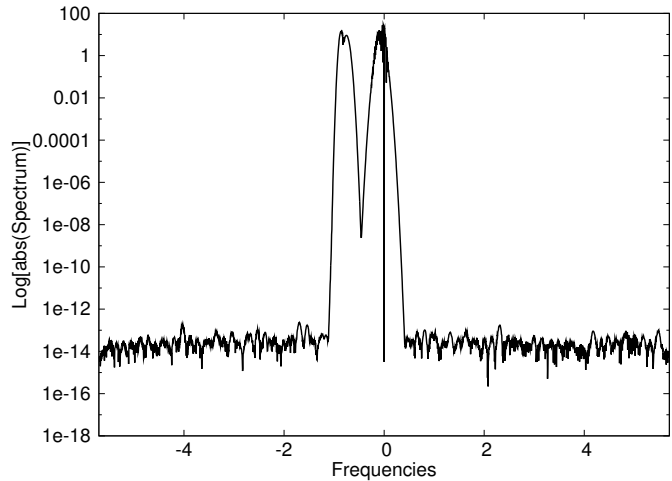}
\caption{Absolute value of the spectrum of $\tilde{f}(t)$ corresponding to \eqref{soustraction} and \eqref{ex2}. The function is given as a sample of $N=2048$ values equally distributed on the interval $[0,T]$.}
\label{SpecEx2}
\end{figure}
In this example, the FFT procedure uses the function $\tilde{f}(t)$ defined by \eqref{soustraction}. This
modified function is characterized by a spectrum with a central component $\ell=0$ equal to zero. This can be seen 
in figure \ref{SpecEx2}, which shows the modulus of the spectrum of $\tilde{f}(t)$. 
As the functions converge rapidly to zero at $t=0$ and at $t=T$, the continuity of the function $\tilde{f}(t)$
(assumed to be a periodic function by the FFT procedure) and also the continuity of the first two derivatives
are ensured without the need 
to extend the time interval or to use an asymptotic polynomial (eq \eqref{polynome}). 
Figure \ref{SpecEx2} confirms
that the functions are well represented by the Fourier expansion; the components omitted in the
spectrum are smaller than 
$10^{-13}$.

\begin{figure}[htp]
\centering
\includegraphics[width=0.7\linewidth]{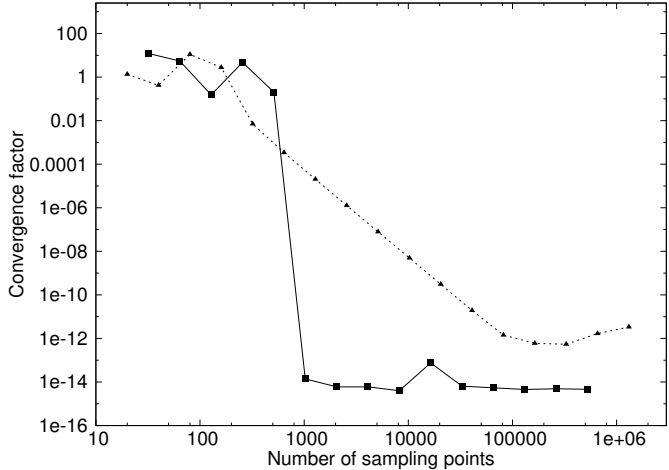}
\caption{Convergence factor \eqref{CF} for the FFT algorithm ($\blacksquare$) and the
Simpson algorithm ($\blacktriangle$).}
\label{CONVEx2}
\end{figure}

\begin{figure}[htp]
\centering
\includegraphics[width=0.7\linewidth]{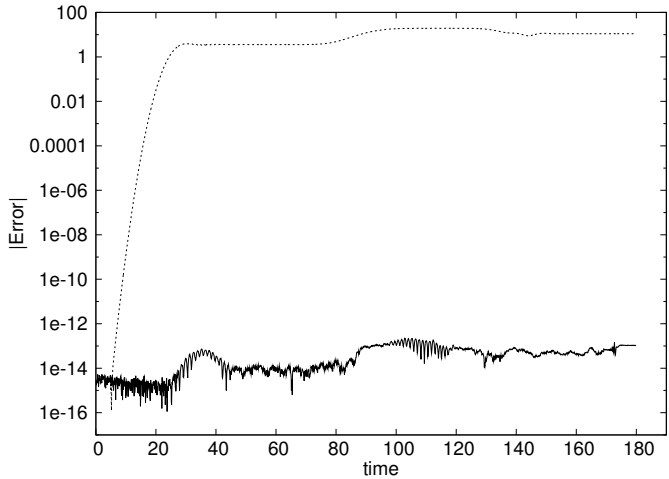}
\caption{Difference between the FFT and the Simpson result: $|I^{FFT}(t_j)-I^{Simp}(t_j)|$ (full line) and
the modulus of the FFT result $|I^{FFT}(t_j)|$ (dashed line).}
\label{PRECIEx2}
\end{figure}

We first compare the speed of convergence of two algorithms, 
the numerical FFT-integrator (this work) and the
standard Simpson rule. 
As the exact values of the integrals $I_j=I(t_j)$ are not known, 
we perform two series of calculations by doubling 
the number $N$ of sampling points, comparing for each of the two integration procedures the result for $2N$ points 
with that for $N$ points. The convergence factor $\mathrm{CF}_N$ is calculated by using
 the $N$ points (with even indices)  that the $2N$ grid has in common with the $N$ grid.
At each of the N sampling points the absolute value of the difference between the two
results for the integral is calculated and the
convergence factor is defined as their maximum value, namely
%
%
\begin{equation}
\mathrm{CF}_N=\max_{j=0}^{N-1}(|I_{2j}^{(2N)}-I_j^{(N)}|).
\label{CF}
\end{equation}
The corresponding numerical results are shown in figure \ref{CONVEx2}. 
As the number of sampling points increases, the FFT convergence factor decreases rapidly by many orders of magnitude up to a
critical $N$ value, 
$N_{FFT}^{cri}=1024$. 
Beyond this point  the convergence factor reaches
a constant plateau with a precision close to
machine precision ($10^{-14}$). 
This is the usual behaviour expected from the Nyquist-Shannon theorem \cite{Shannon} when using Fourier basis expansions. 
The Simpson rule results show a strongly different behaviour. 
The convergence factor decreases regularly up to a minimum value for a critical value
$N_{Simp}^{cri}$, which is about 100 times larger than for the FFT case. Moreover
the minimum convergence factor reached is about two orders
of magnitude greater than for the FFT results. 
After this minimum value, the convergence factor increases again. 
In this new regime the roundoff errors become predominant. 
Since no exact result is available, 
we next compare the results obtained with the best precision for each method, to
be sure that the two algorithms are consistent and converge to the same value. 
This corresponds to $N=1024$ for the FFT algorithm and to about $N=150000$ for the Simpson algorithm.
Figure \ref{PRECIEx2} presents for each discrete time value the modulus 
of the difference between the
two integrals calculated with the FFT algorithm and with the Simpson rule,
\begin{equation}
|I^{FFT}(t_j)-I^{Simp}(t_j)|.
\end{equation}
This figure reveals that the difference on the whole 
interval $[0,T]$ is never larger than $10^{-13}$. Similar results, not presented here, are obtained 
for examples in which it was necessary to impose a continuity of the two first derivatives 
by using a $6^{th}$ order polynomial (see \ref{int_non_per}).
These applications prove that the FFT integrator possesses the expected qualities.
Figure \ref{PRECIEx2} shows
that the precision of the results is very high 
and virtually independent of the discrete instant $t_j$ taken as the upper bound. 
Moreover, figure \ref{CONVEx2}
reveals that the number of sampling points necessary to give convergence for the FFT is 100 times less than that for the
convergence of the Simpson algorithm. Finally the operations necessary for the FFT algorithm are reduced to
the calculation of two FFT, the first being made in any case during the propagation scheme \cite{kosloff1996}.
For the selected example the CPU time associated with the FFT procedure is about 20 times smaller than that associated with the Simpson rule. 
However, the degree of advantage of the FFT approach is 
closely related to the shape of the spectrum of $f(t)$. In the present cases as well as for the
Schr\"odinger equation integration procedure treated in the next section, the spectrums are narrow and are thus 
well represented by a Fourier basis set.

\section{An asymmetric double-well transition experiment \label{results}}
\begin{figure}[htp]
\centering
\includegraphics[width=0.7\linewidth]{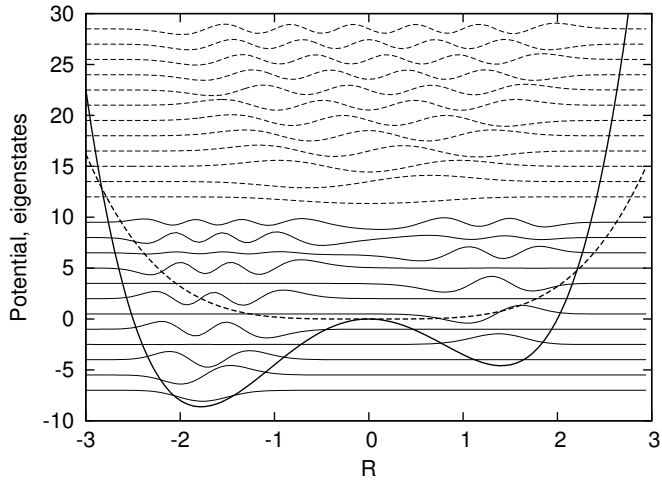}
\caption{The two potential energy curves and some of the first unperturbed vibrational eigenstates
of the first potential surface (full lines) and of the second one (dashed lines).
The potentials are quartic polynomials defined as: 
$\epsilon_1(R)= -5 R^2 + 0.5 R^3 + R^4 $ ; 
$\epsilon_2(R)= 0.2 R^4$.
}
\label{Potential}
\end{figure}

The illustrative example is that of a model  
diatomic molecule submitted to two laser pulses. 
Throughout this section we use arbitrary units with the convention $\hbar =1$. 
The various numerical parameters have been adjusted so that the relative orders of magnitude of
the vibrational level spacing, the strength of the dipole couplings and the pulse duration
produce realistic dynamics with significant transition probabilities. 
%
Two potential curves 
$S=1,2$ 
are defined, which can refer to 
the two first electronic states of a one-dimensional vibrational Hamiltonian in
the framework of the Born-Oppenheimer approximation. 
These two curves 
are shown in figure \ref{Potential}.
The lower surface is a double well similar to those used in references 
\cite{Bavli1993,Wadehra2003},
\begin{equation}
\epsilon_1(R)= -5 R^2 + 0.5 R^3 + R^4.
\end{equation}
The upper surface is a single quartic well,
\begin{equation}
\epsilon_2(R)= 0.2 R^4.
\end{equation}
The $R$ coordinate is chosen as a relative position centred on the barrier of $\epsilon_1$. 
All the eigenstates are pure bound states
and were calculated using a Fourier grid method on the radial coordinate 
\cite{Marston1989,Kosloff1983}. They are shown in figure \ref{Potential}.
These model potentials are also similar to those used to describe, for instance, effective isomerization problems along a reaction coordinate \cite{chenel2012}. 
The evolution of the vibrational 
evolution operator is driven by the following equation within
the framework of the dipole approximation:
%
%
\begin{equation}
i \frac{\partial}{\partial t} U(t,0)=\left[ T_N +\left(
\begin{array}{cc}
\epsilon_1 & -\vec{\mu}_{1,2}.\vec{E}(t) \\
 -\vec{\mu}_{1,2}.\vec{E}(t)& \epsilon_2 
\end{array}\right) \right] U(t,0)
\label{dyn1}
\end{equation} 
%
$T_N$ is the relative kinetic energy of the two atoms, i.e. $T_N = -\frac{1}{2m}\frac{\partial ^2}{\partial R^2}$
with an arbitrary mass equal to $10$.  
$\epsilon_{S=1,2}$ are the two energy surfaces. 
There is no rotation in the model and the dipole transition moment 
${\vec{\mu}}$
is assumed to be constant 
along the radial axis $R$ and equal to $1$. 

\begin{figure}[htp]
\centering
\includegraphics[width=0.45\linewidth]{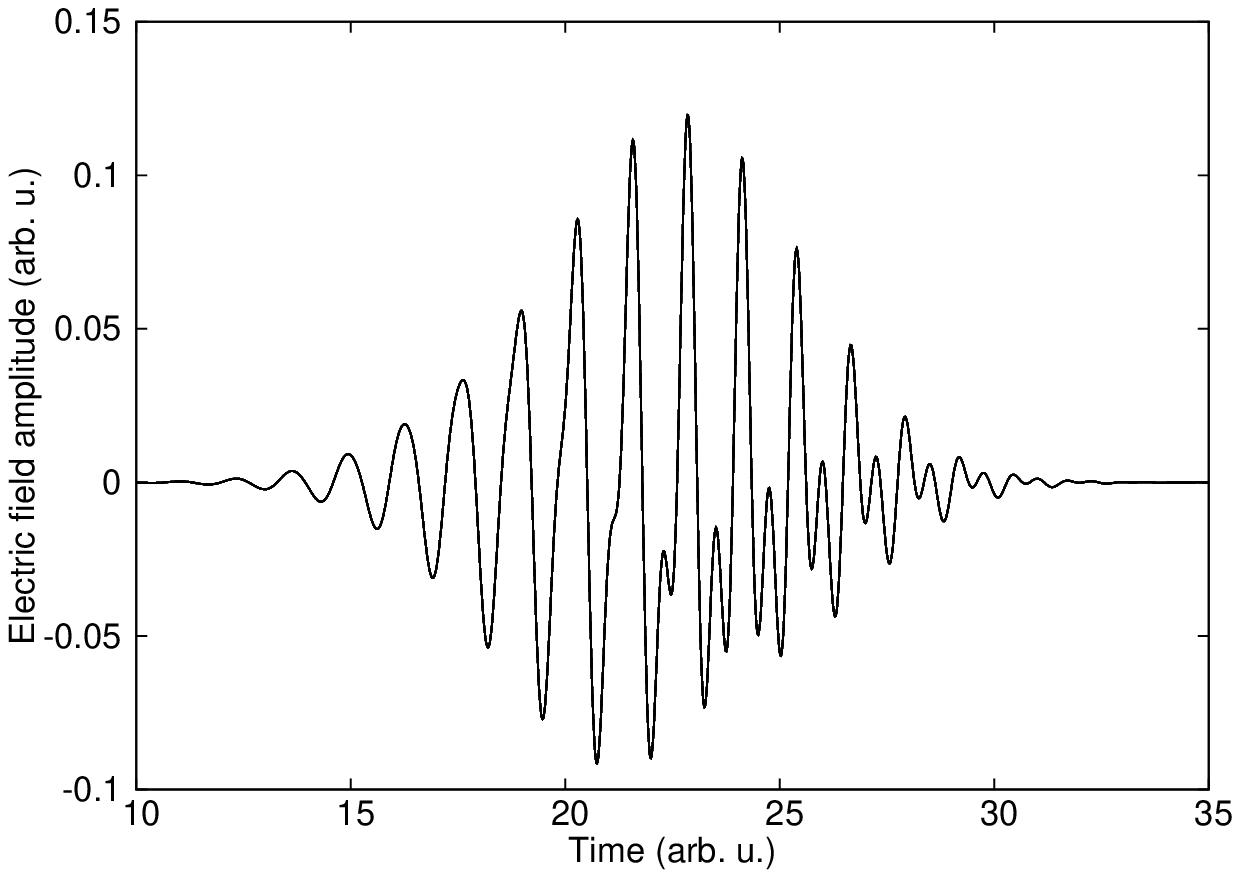}
\includegraphics[width=0.45\linewidth]{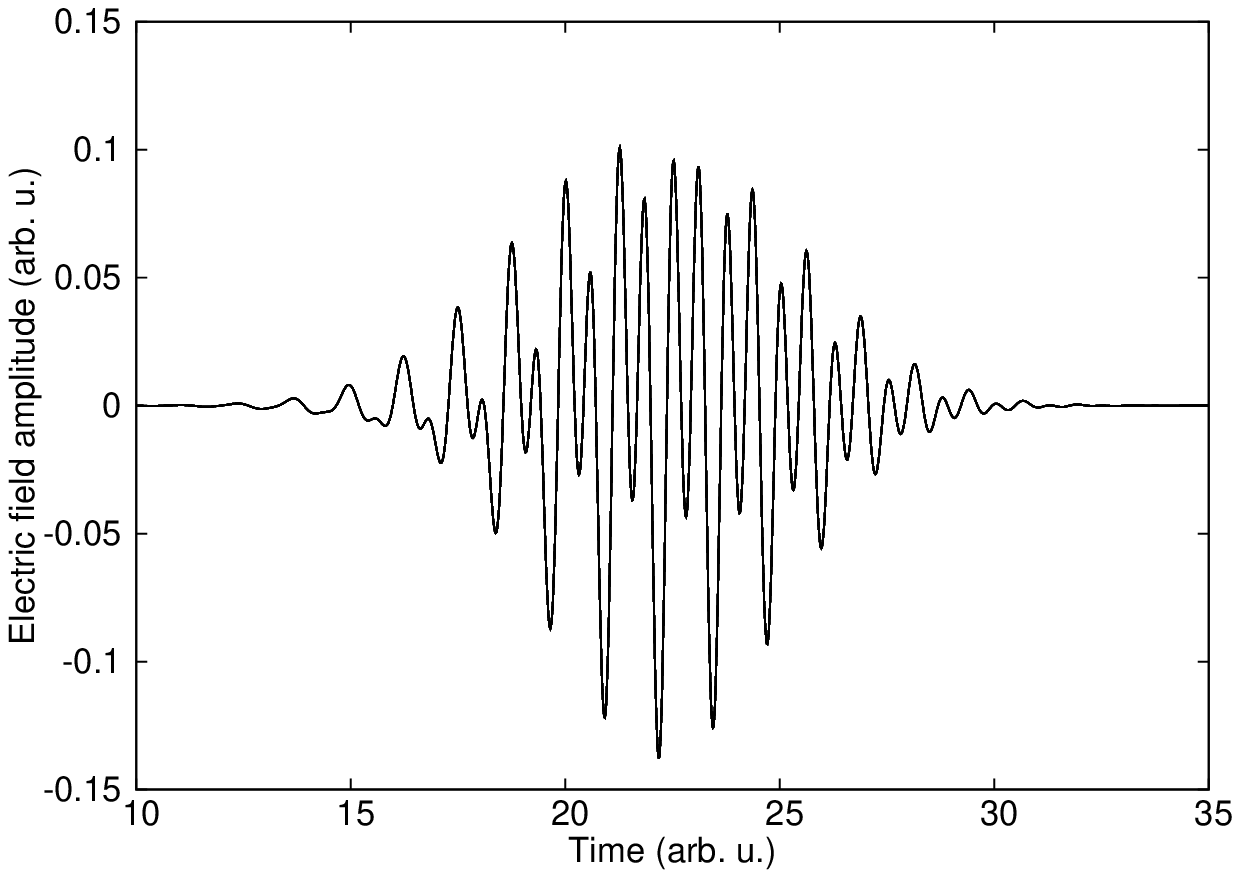}
\caption{The electric field $E(t)$ in the first example of \eqref{parchamp1} (left frame) and in the second example of \eqref{parchamp2} (right frame).}
\label{Efield1}
\end{figure}

The laser field  amplitude projected on the ${\vec{\mu}}$ direction
is defined as 
the sum of two 
pulses with gaussian envelopes
%
%
\begin{equation}
E(t)=\sum_{j=1}^ 2 E_j \; \cos \left(\omega_j(t-T_j) \right) \; \exp \left(-\left(\frac{t-T_j}{\tau_j} \right)^2\right)
\label{cham}
\end{equation}
%
Vibrational states $(v=1,S=1)$ and $(v=6,S=1)$ have in common a strong overlap with state $(v=7,S=2)$ of the excited  potential curve. We have therefore chosen the carrier frequencies as $\omega_1= E(v=7,S=2)-E(v=1,S=1) \simeq 9.98449$ and $\omega_2=E(v=7,S=2)-E(v=6,S=1) \simeq 4.77725$. 
This choice induces resonant transitions between the first 
and the sixth eigenstate of the lower surface: 
$\omega_1 -\omega_ 2=E(v=6,S=1)-E(v=1,S=1)$. 
This will ensure strong transitions between the two asymmetric wells and a rather large depletion of the initial state. 
Calculations were made using $\tau_1=3.90$ and $\tau_2=4.50$ and two different electric field amplitudes and centres,
\begin{eqnarray}
E_1=0.05, \; T_1= 23.5 \nonumber \\
E_2=0.08, \; T_2= 21.5
\label{parchamp1}
\end{eqnarray}
in the first example and 
\begin{eqnarray}
E_1=0.09, \; T_1= 22.5 \nonumber \\
E_2=0.05, \; T_2= 21.5
\label{parchamp2}
\end{eqnarray}
in the second example. 
The corresponding 
electric fields are represented in figure \ref{Efield1}.
Those choices for the laser parameters place the system in a non optimal STIRAP configuration
\cite{vitanov2001}. 
In both
cases the initial state is the first eigenstate of the first surface $(v=1,S=1)$. 
In terms of the wave operator, the first global iteration starts with a guess chosen as
$\Omega^{(0)}(t)=\ket{ v=1,S=1 } \langle v=1,S=1 \vert$, i.e. $X^{(0)}(t)=0$. 
A vibrational basis set made of 
the 30 first vibrational eigenvectors of each surface 
(i.e. a total molecular basis set of 
 $N_v =60 $ states) 
is sufficient to obtain converged results. 
This molecular basis set is small but here the principal aim is to test the time propagation algorithm. 
Here we need $N_t= 4096$ Fourier states $|p\rangle$
with $\langle t|p\rangle =\exp (i 2\pi p t/T)$
(or equivalently $N_t=4096$ time grid points) 
to describe 
the time evolution correctly. 
In what follows 
the vibrational states are numbered from $v=1$ to $v=30$ for the first surface and
from $v=31$ to $v=60$ for the second surface. 
The total basis set describing the extended Hilbert space is then composed of 
$N=N_v \times N_t=245760$ states. 
In this framework, the dynamics is calculated 
by using the iterative scheme 
explained in section \ref{iterative_proc} 
(equations \eqref{TOTAL}, \eqref{Heff}, \eqref{SOLRIG1} and \eqref{SOLRIG} with $\Omega^{(n+1)}=P_o+X^{(n+1)}$
and $X^{(n+1)}=X^{(n)}+\delta X^{(n)}$.)
The integrals appearing in \eqref{SOLRIG}
are calculated using 
the numerical integrator presented in section \ref{integrals} and \ref{app_FFT}. 
This scheme
is used to calculate  all the components of the off-diagonal part of the wave operator for each of the 
$N_t$ discrete $t_k$ values: $X^{(n)}_{v}(t_k)$, $v=1,\dots,N_v$, $k=1,\dots,N_t$, at each iteration order $(n)$. 

\begin{figure}[htp]
\centering
\includegraphics[width=0.7\linewidth]{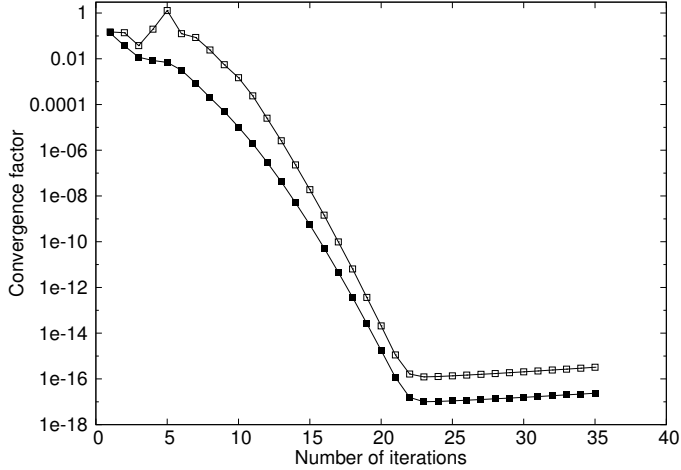}
\caption{Convergence factor \eqref{FC} with the parameters defined in \eqref{parchamp1} ($\blacksquare$)
and those defined in \ref{parchamp2} ($\square$) with respect to the iteration number.} 
\label{SPEED}
\end{figure}
\begin{figure}[htp]
\centering
\includegraphics[width=0.7\linewidth]{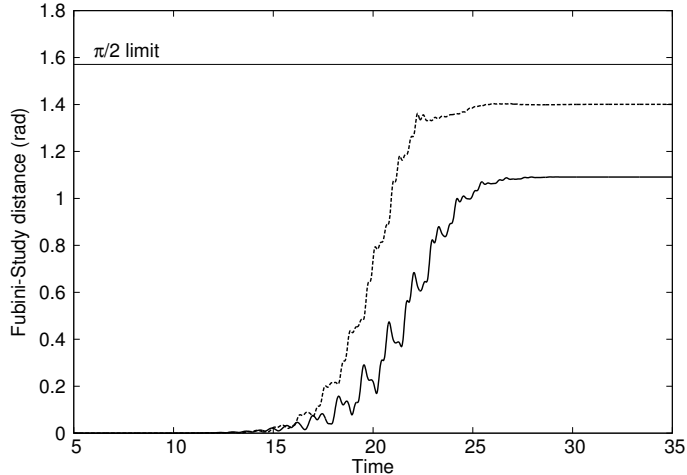}
\caption{Evolution of the Fubini-Study distance $dist_{FS}(P_o,P(t))$ during the pulse
defined in \eqref{parchamp1} (solid line) and \eqref{parchamp2} (dashed line).
The distance has been computed using the wave operator at iteration 22.} 
\label{fubini}
\end{figure}

Our main concern is the speed of convergence of the global iterative scheme. 
We define the following convergence factor:
%
%
%
\begin{equation}
  F=\frac{\sum_{v =1}^{N_v}\sum_{k=1}^{N_t}|\delta X^{(n)}_{v}(t_k)|^2}
  {\sum_{v =1}^{N_v}\sum_{k=1}^{N_t}| X^{(n+1)}_{v}(t_k)|^2},
\label{FC}
\end{equation}
with $\delta X^{(n)}$ defined in \eqref{SOLRIG}. 
Figure \ref{SPEED} shows that
the two calculations converge. 
A plateau of about $10^{-17}$ for the first example and of about $10^{-16}$
for the second one is reached at the $22^{th}$ iteration. 
Between the $10^{th}$ and the $22^{th}$  iteration,
the accuracy is improved by one order of magnitude at each iteration. 
The convergence factor of equation \eqref{FC} 
adds the residues $\delta X^{(n)}(t)$ over the whole propagation interval.
This means that the plateau seen in figure \ref{SPEED} indicate a very high accuracy at each time grid point. 
It is essential to note that, by contrast with a continuous
propagation scheme, 
in which errors are accumulated at each time step, the global scheme does not accumulate the errors between two iteration orders. 
Nevertheless our iterative procedure possesses, as does every iterative treatment, 
a finite radius of convergence. 
The small increase of the convergence factor observed in figure \ref{SPEED} during 
the first four iterations for the second 
example is 
the consequence of a large Fubini-Study distance between $P_o$ and $P$ in this case 
(the survival probability at the end of the interaction $|\langle v=1|\Psi(t_{final})\rangle|^2$ is 0.2129
in the first example but only 0.02864 in the second one).
This increases the distance between $P_o$
and $P(t_{final})$ in this last case.   
%
This can be seen in figure \ref{fubini} which illustrates the progressive separation 
between the active subspace $S_o$ and the dynamic subspace $S(t)$. 
In the non-degenerate case, the Fubini-Study distance is simply
\begin{equation}
dist_{FS}(P_o,P(t))= \arccos \left( \parallel \langle v_i \vert \Psi (t) \rangle \parallel \right) = \arccos \left( \frac{1}{\parallel \Omega (t) \parallel} \right).
\end{equation}
The right-hand side is obtained using the normalization properties of the wave operator. 
The final distance for the second example of \eqref{parchamp2} is clearly larger. 
The physical reason explaining this strong difference is that example \eqref{parchamp2} is closer to an efficient STIRAP situation than example \eqref{parchamp1} (but still non-optimal, the final transition probability to state $\vert v=6 \rangle$ being here only 0.22, because of many non-adiabatic transitions to other vibrational states). 

\begin{figure}[htp]
\centering
\includegraphics[width=0.7\linewidth]{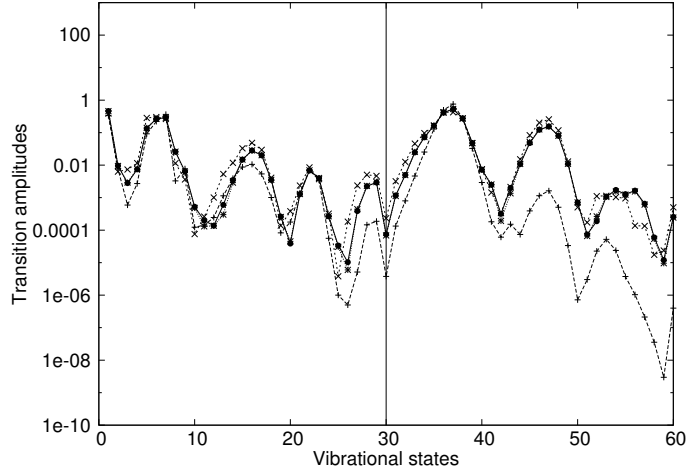}
\caption{Final transition amplitudes to the different vibrational states $|\langle v| \Psi^{(n)}(t_{final})\rangle|$ after the pulse defined in  \eqref{parchamp1}, 
obtained from the wave operator at successive iteration orders. $+ : (n=2), \times : (n=4), \ast : (n=9), \bullet : (n=22)$.} 
\label{TP1}
\end{figure}

\begin{figure}[htp]
\centering
\includegraphics[width=0.7\linewidth]{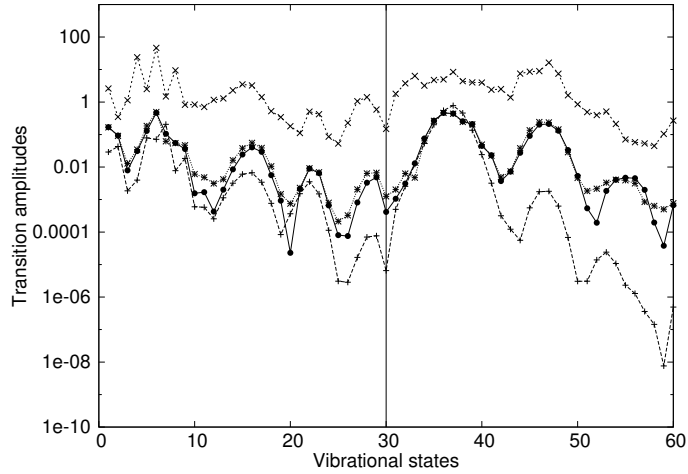}
\caption{Final transition amplitudes to the different vibrational states $|\langle v| \Psi^{(n)}(t_{final})\rangle|$ after the pulse defined in  \eqref{parchamp2}, 
at successive iteration orders. $+ : (n=2), \times : (n=4), \ast : (n=9), \bullet : (n=22)$.} 
\label{TP2}
\end{figure}

\begin{figure}[htp]
\centering
\includegraphics[width=0.7\linewidth]{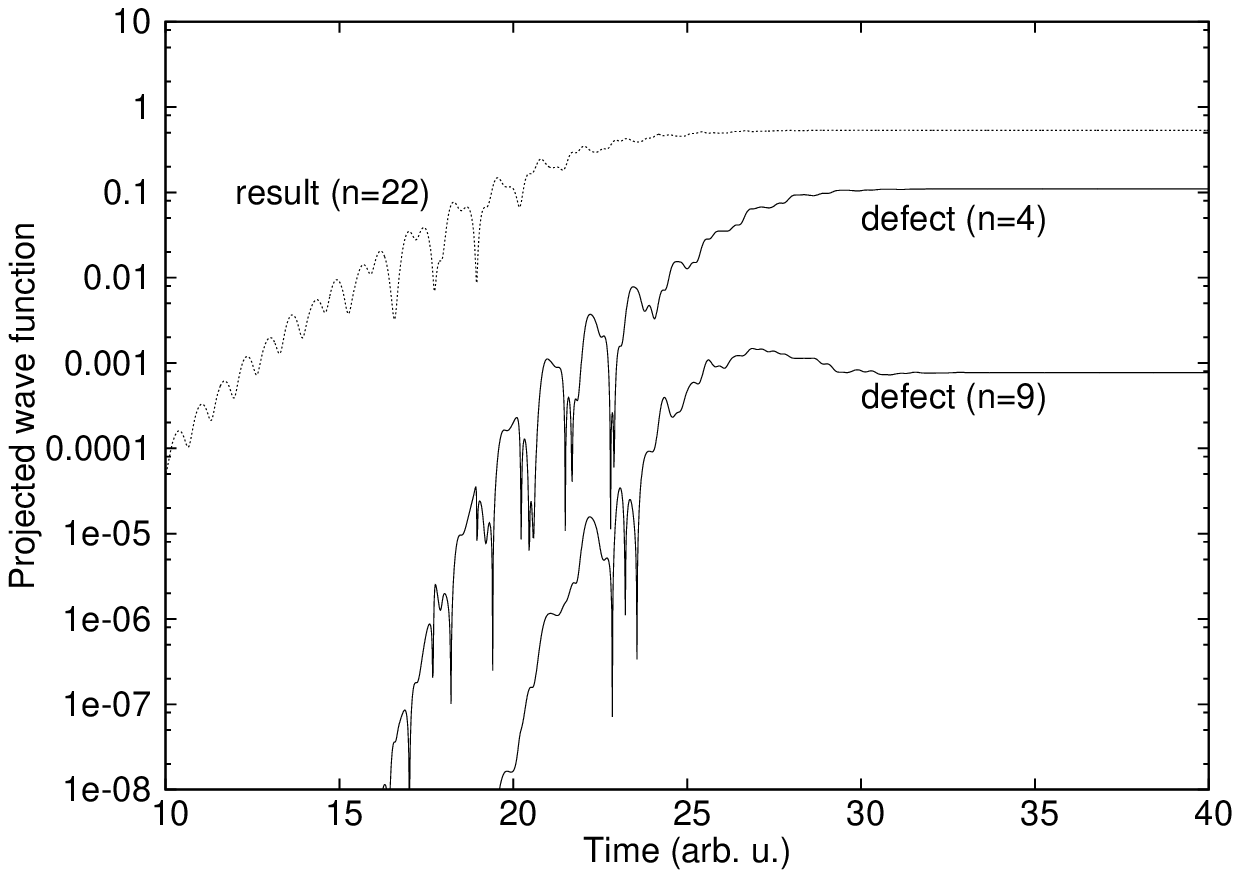}
\caption{The dashed line represents the modulus of the projected wave function $|\langle v=37|\Psi^{(n)}(t)\rangle|$
as a function of time 
at the end of the iterative process $(n=22)$ for the first example (parameters given in \eqref{parchamp1}). 
The full lines are the defects 
at intermediate iteration orders: 
$|\langle v=37|\Psi^{(n=22)}(t)-\Psi^{(n)}(t)\rangle | $ for $(n=4)$ and $(n=9)$.}
\label{TP1T}
\end{figure}

\begin{figure}[htp]
\centering
\includegraphics[width=0.7\linewidth]{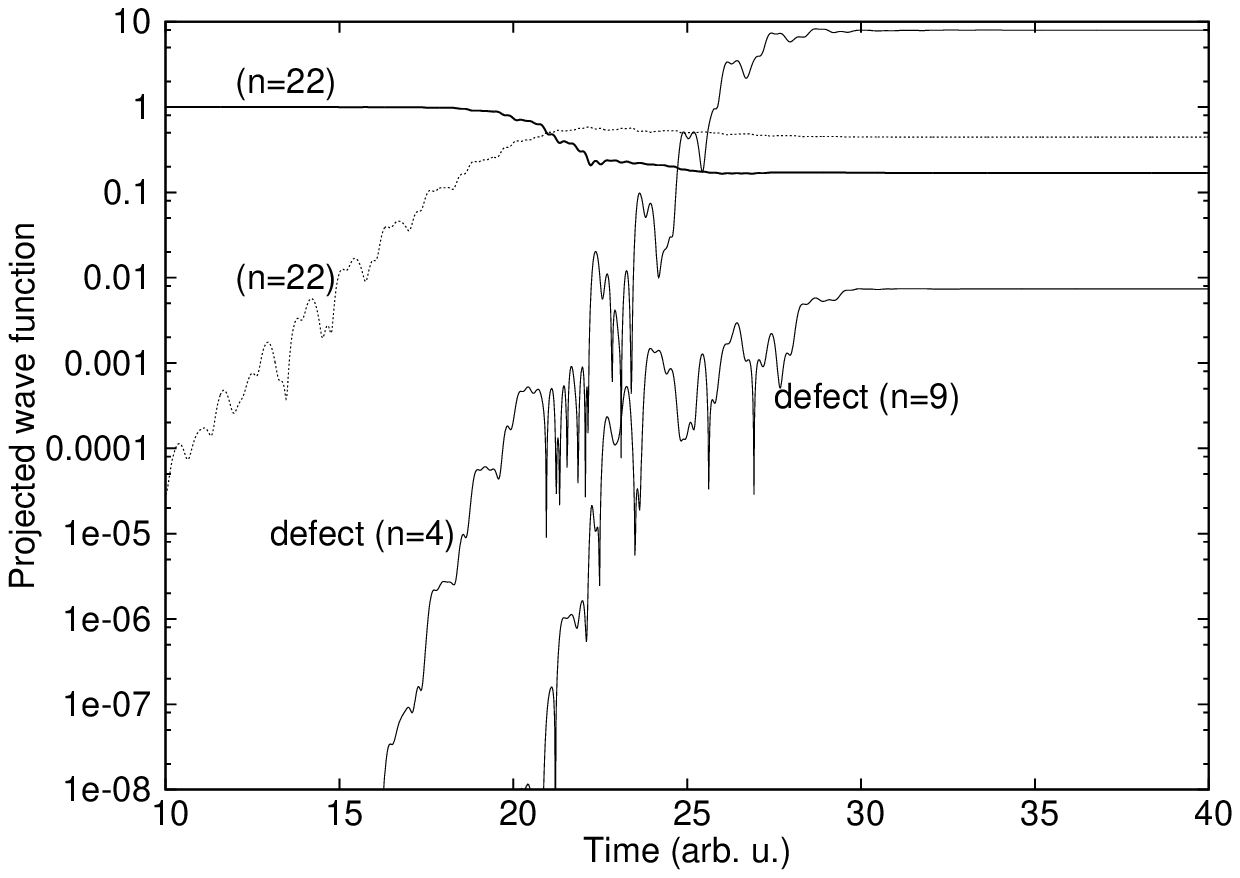}
\caption{The dashed line represents the modulus of the projected wave function $|\langle v=37|\Psi^{(n)}(t)\rangle|$
as a function of time 
at the end of the iterative process $(n=22)$ for the second example (parameters given in \eqref{parchamp2}). 
The thin continuous lines are the defects at intermediate iteration orders :
$|\langle v=37|\Psi^{(n=22)}(t)-\Psi^{(n)}(t)\rangle | $ for $(n=4)$ and $(n=9)$. 
The deep black line shows
the survival amplitude 
$|\langle v=1|\Psi^{(n=22)}(t)\rangle|$.}
\label{TP2T}
\end{figure}

The global character of the integration is evident in figures \ref{TP1} and \ref{TP2} which present the transition
moduli to the different vibrational states at the final time 
$|\langle v| \Psi^{(n)}(t_{final})\rangle|$
at four different iteration orders. 
It is interesting to note that in figure \ref{TP1}, an approximate vibrational
distribution close to the exact one is obtained 
as early as at the fourth iteration. 
The small convergence difficulties 
during the first iteration orders 
in the second example are evident in figure \ref{TP2}. 
The $4^{th}$ order shows some transition amplitude 
larger than 50, very far from the final results. 
In this case more iterations are needed to achieve a stabilized distribution. 
%
Probabilities can thus exceed one in the global scheme if the results are observed at an early intermediate order, before the completion of the iterative process, such as in Fig. \ref{TP2}. There is no fixed norm conservation in the structure of the algorithm but the norm conservation is recovered after convergence. 
%
%
Figures \ref{TP1T} and \ref{TP2T} also illustrate the global character of the time-dependent propagation. 
In these figures we present the evolution of the transition amplitude to state $|v=37\rangle$. 
This particular state is strongly coupled with the initial state $\vert v=1 \rangle$. 
Figure \ref{TP1T} shows that each iteration order produces a projected amplitude which converges
towards the exact solution 
over the whole interaction time interval. 
At the fourth iteration, the difference $|\langle v=37|\Psi^{(n=22)}(t)-\Psi^{(n)}(t)\rangle |$ is 
already lower than $10^{-1}$. 
At the $9^{th}$ order, the same error 
becomes lower 
than $10^{-3}$. 
Figure \ref{TP2T} reveals the difficulties due to the large Fubini-Study
distance between $P_o$ and $P(t_{final})$
(here the final survival probability is $|\langle v=1|\Psi(t_{final}\rangle|^2= 0.02864$). 
As a consequence, the amplitude of the projected wave function onto the state
$v=37$ at the $4^{th}$ iteration order is about 10 times larger than the exact one for $t>30$. 
Surprisingly, 
the iterative procedure converges again 
between the $4^{th}$ and the $9^{th}$ order. 
At the $9^{th}$ order
the error is smaller than $10^{-2}$ everywhere. 

We think that this efficiency is essentially due to the use
of a time-dependent effective Hamiltonian in the basic equations
of the integration scheme (\eqref{Heff},  \eqref{SOLRIG1} and \eqref{SOLRIG}), 
especially when it is compared with previous iterative schemes. 
\begin{figure}[htp]
\centering
\includegraphics[width=0.7\linewidth]{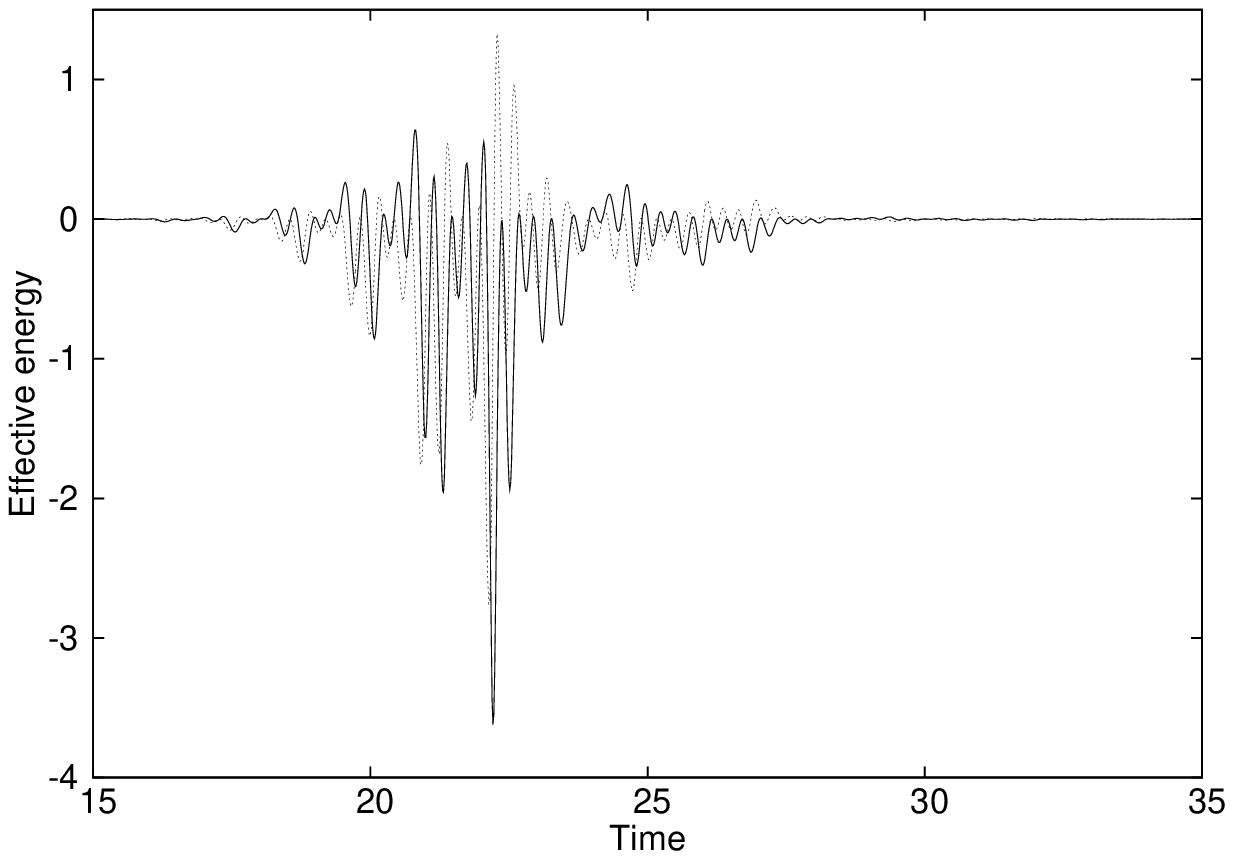}
\caption{Evolution of the real part (continuous line) and imaginary part (dashed line) of the effective energy, using the field parameters  
\eqref{parchamp2}.} 
\label{HEFF2}
\end{figure}
Figure \ref{HEFF2} 
reveals large oscillation of the effective Hamiltonian defined in \eqref{TOTAL}. 
The imaginary part imposes the unitarity 
of the evolution in the wave operator formulation. 
A contribution to the evolution operator of the form
$$\exp \left(\frac{1}{i\hbar}\int_0^{t_{final}} \Im m (H_{eff} )dt'éé \right) $$
is present and gives the correct survival amplitude. 
The magnitude of 
these oscillations ($ \approx 4$ for the real part) 
is even much larger than the distance between the
first 
closest vibrational eigenvalues of each surface (for example $E(v=2)-E(v=1)=1.451$). 
Other iterative methods in which these time variations are neglected completely fail to converge on this same numerical example. 
For example, 
the adiabatic approximation would be justified if in equation \eqref{SOLRIG} the amplitude
$|\langle v|\tilde{H}^{(n)}_{diag}(t)|v\rangle-\langle v_i|H^{(n)}_{eff}(t)|v_i\rangle|$
is large at each instant, $\vert v_i \rangle$ being the initial vibrational state. 
In such a case the introduction of the approximation \eqref{ADAP} into equation \eqref{SOLRIG}
leads to: 
%
\begin{equation}
\delta X^{(n)}(t)=\frac{i\hbar \Delta^{(n)}(t)}{\tilde{H}_{diag}^{(n)}(t)-H^{(n)}_{eff}(t)}
\label{ADEQ}
\end{equation}
Equation \eqref{ADEQ} is very similar to the RDWA iterative method which has already been used to compute time-dependent wavepacket propagations \cite{Jol3,Lecl3}. 
However it cannot be used in the present case. 
It is evident that the strong time-dependency of $H_{eff}(t)$ (see figure \ref{HEFF2}) and the very small survival probability at $t_{final}$ are inconsistent with the use of any adiabatic approximation. 
Actually the use of \eqref{ADEQ} in the integration scheme produces a very 
fast divergence of the iterative series caused by the crossings of $\tilde{H}^{(n)}_{diag}$ and $H^{(n)}_{eff}$. 
Another approximate (but more sophisticated) integration scheme was proposed in refs \cite{Jol2} and \cite{Jol3}. 
By using a discrete Fourier basis set: $ |p\rangle$ with $\langle t|p\rangle =\exp (i 2\pi p t/T)$ derived from the discrete time grid basis set by a FFT transformation, 
a generalised RDWA approximation leads to: 
%
%
\begin{eqnarray}
\langle v',p'|X^{(n+1)}|v_i,p=0\rangle \nonumber \\
= \frac{\langle v',p'| \left[ (H_F-\tilde{H}^{(n)})X^{(n)}+H  \right] |v_i,p=0\rangle}
{\langle v_i,p=0|H_{eff}^{(n)}|v_i,p=0\rangle-\langle v',p'|\tilde{H}^{(n)}|v',p'\rangle}.
\label{ADEQT}
\end{eqnarray} 
%
Introducing \eqref{ADEQT} into the iterative scheme also produces a divergence, albeit 
slower than the one produced by \eqref{ADEQ}. 
Undoubtedly this defect comes from the fact that equation \eqref{ADEQT} only uses an averaged $H_{eff}$, 
represented by the first Brillouin component $\langle p=0|H_{eff}|p=0\rangle$, rather than the exact and strongly time-dependent expression. 


Finally it is crucial to verify that the global scheme gives the correct results
by a comparison with some continuous wave-packet propagation algorithms. 
We first compare our results with those obtained for the same system by 
using a split operator \cite{feit} procedure
by separating the role of the diagonal unperturbed Hamiltonian and the role of the laser-molecule coupling. 
This last contribution is not diagonal and its role is approximated by using a Second Order Differencing (SOD) scheme. 
We also calculate the same dynamics using the short iterative Lanczos (SIL) method \cite{park}. 
To be significant, the same vibrational basis set of $N_v=60$ states is used in the continuous propagations
(for the SIL propagations, the size of the Lanczos subspace was set up to 10). 
\begin{figure}[htp]
\centering
\includegraphics[width=0.7\linewidth]{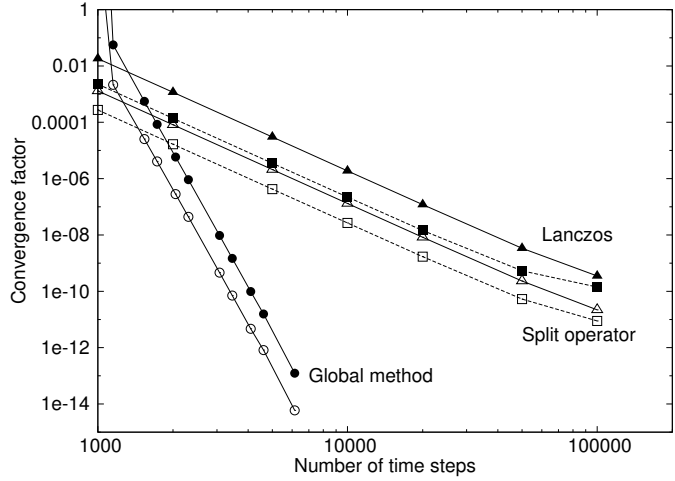}
\caption 
{Convergence factors $\mathcal{F}_C$ and $\mathcal{F}_G$ defined in equations \eqref{FC12} and \eqref{FG12}
for the global method (circles), the short iterative Lanczos scheme (triangles) and the split-operator algorithm (squares)
using the field parameters of equation \eqref{parchamp1} 
($\square$, $\triangle$ and $\circ$)
and of equation \eqref{parchamp2}
($\blacksquare$, $\blacktriangle$ and $\bullet$),
as a function of the number of time steps.}
\label{COMP12}
\end{figure}
%
%
A convergence factor is defined as
%
%
\begin{equation}
\mathcal{F}_C(N_t)=\sum_{i=1}^{N_v} |\Omega^{C,N_t}(i,t_{final})-\Omega^{G,4096}(i,t_{final})|^2.
\label{FC12}
\end{equation} 
$\Omega^{G,4096}$ is the wave operator calculated at the final time 
using the global propagator
with $4096$ time grid points 
after 22 iterations and $\Omega^{C}$ is the wave operator reconstructed 
by using equation \eqref{WOD} after a complete propagation with the continuous propapagators
with $N$ time steps.  
To show the convergence of the global method itself we have also computed a similar convergence factor between global results, 
\begin{equation}
\mathcal{F}_G(N_t)=\sum_{i=1}^{N_v} |\Omega^{G,N_t}(i,t_{final})-\Omega^{G,8192}(i,t_{final})|^2. 
\label{FG12}
\end{equation} 
Figure \ref{COMP12} shows the evolution of $\mathcal{F}_C$ and $\mathcal{F}_G$ with respect to the number of discrete time steps. 
This figure reveals that the three procedures converge to the same final values (same amplitudes and same phases). 
Figure \ref{COMP12} also confirms that the number of time steps needed to converge to a given accuracy is much smaller with the global scheme than with the continuous schemes.
The convergence is faster for the first pulse than for the second one.  
The number of time-steps of length $dt$ for the continuous integrator is quite large especially if a high accuracy is needed. 
The CPU time needed to compute $\delta X^{(n)}$ (equation \eqref{SOLRIG}) at 
a given iteration order is of course much larger than that required to construct one step in the continuous
 algorithms. 
Nevertheless the convergence plateau 
in figure \ref{SPEED} are reached after only 22 iterations. 
With the field parameters of equation \eqref{parchamp1}, passing from iteration 22 to iteration 23 indicates that 
the transition amplitudes are converged with 7 stable and significant digits. 
%
%
%
More detailed informations about the CPU time are given in table \ref{tpsCPU} with varying required accuracies. 
This table shows the ratio between the CPU times required to reach a given threshold of convergence using the global or the continuous strategies, using the convergence factors $\mathcal{F}_G$ and $\mathcal{F}_C$ defined in \eqref{FC12} and \eqref{FG12}.  
The results of the global method have been observed after 22 iterations. 
\begin{table}[htp]
\centering
\begin{tabular}{lllllll}
\hline
\hline
Convergence factor $<$ 				& $10^{-6}$ & $10^{-8}$ & $10^{-10}$ & $10^{-12}$ & $10^{-14}$ \\
\hline
Stable digits $\simeq$ 				& 3 	& 4 	& 5 	& 6 	& 7 \\
$T^{CPU}_G / T^{CPU}_{SIL} $ 		&  3.83 	& 3.11 	& 0.775	& 0.757	& 0.211 \\
$T^{CPU}_G / T^{CPU}_{Split-Op.}$ 	&  15.3 & 6.52 & 3.31 & 1.62 & 0.912 \\
\hline
\end{tabular}
\caption{Approximative ratios of the CPU times required to reach a given convergence threshold between the global (G) and the continuous methods (SIL or split-operator), for example of equation \eqref{parchamp1}. The corresponding number of significant digits on the largest transition probabilities are also given. }
\label{tpsCPU}
\end{table}
The continuous algorithms are faster if a low accuracy is sufficient. 
The SIL algorithm and the global method then give comparable CPU times for intermediate accuracies,
while the split operator stays a little faster. The global method becomes slightly faster if a high accuracy is desired. 

\section{Conclusion \label{conclusion}}

The theoretical analysis and the simple illustrative examples presented in this paper
demonstrate that a short global iterative scheme can solve the time-dependent Schr\"odinger 
equation with an explicitly time-dependent Hamiltonian operator.  
This global method is competitive with a continuous step by step algorithm. 
The following points should be highlighted: 

i) 
A global integrator does not accumulate discretization errors
along the time interval and consequently can generate high precision results. 
The numerical experiments have shown that each new iteration order increases the precision by one order of magnitude. 
The structure of the algorithm also allows for a rigourous estimation of the error associated with each iteration order by following the quantity $H_{F} \Omega - \Omega H \Omega$, the calculation being stopped when the expected accuracy is reached.  

ii) The use of an FFT procedure to calculate the numerous time integrals 
appears as a very 
efficient choice which is compatible with producing 
a highly accurate solution to the Schr\"odinger equation.
Nevertheless it is essential that the Fourier spectrum of the perturbation should be well reproduced by the Fourier basis set
and it is certainly more difficult to solve problems with broader spectra. 

iii) The global iterative scheme does not introduce limitations concerning the nature or the structure of the
Hamiltonian which drives the dynamics. 
Future applications 
could usefully be aimed at checking how well it can 
reproduce dissipative dynamics. 

iv) The inability of previous global iterative techniques to determine the time-dependent 
wave-operator for similar problems is partly due to a wrong estimation 
of the effective Hamiltonian 
acting in the selected model space. 
In the selected examples $H_{eff}(t)$ exhibits large oscillations 
which must correctly be taken into account. 

v) In the present applications the starting input for the wave-operator 
is simply $X^{(n=1)}=0$. 
The iterative nature of the method makes possible 
a rapid generation of new solutions, after a slight modification of the Hamiltonian parameters, 
by simply starting with the wave-operator obtained from the previous solution. 
Many applications are possible, 
among which we can mention the study of laser control processes using exceptional points of parametric Hamiltonians \cite{EP1,EP2,EP3,EP4} 
or of vibrational cooling using zero-width resonances \cite{atabek2013}.

The algorithm proposed in this paper appears to be useful for investigating new formulations of quantum control in 
molecular physics or for treating complicated near-adiabatic evolutions, 
for which the use of an effective Hamiltonian can efficiently reduce the dimension of the Hilbert space \cite{RUTH}.
The global algorithm add the time as a new quantum coordinate, which can be a disadvantage to compute the dynamics of high dimensional problems over long times. On the other hand the method is not dependent on the structure of the Hamiltonian 
as it can be the case for the split operator algorithm or Magnus expansions. 
The wave operator approach is in principle transposable to dynamical processes like dissociation, ionization, dynamics involving degenerate eigenvalues or non-adiabatic interactions. 
For dissipative processes, 
a special difficulty can appear in situations for which the dynamics escapes too far from the selected model subspace. 
In such cases, the algorithm can diverge. 
The best option is to generalise the iterative algorithm of section \ref{iterative_proc} 
to handle a degenerate active space, by incorporating all the states which are strongly coupled into the model space. 
Study of this generalisation is in progress, as the adaptation of the algorithm to the calculation of Floquet eigenstates. 

\ack
We thank John P. Killingbeck for fruitful discussions and for a careful reading of the English manuscript. 
Simulations have been executed on computers of the Utinam Institute of the
Universit\'e de Franche-Comt\'e, supported by the R\'egion de Franche-Comt\'e and Institut des Sciences de l'Univers (INSU).


\appendix

\section{Numerical FFT integration \label{app_FFT}}

\subsection{Integration of periodic functions}

Consider equation \eqref{solution_continue},
\begin{equation}
I(t)  
=\frac{1}{2}F(0) - \frac{i}{2\pi} PV\int_{-\infty}^{+\infty}\frac{F(\nu)e^{2i\pi\nu t}}{\nu} d\nu,
\end{equation}
where $F(\nu)$ is the Fourier transform of $f(t)$. 
If the function $f(t)$ is assumed to be non-zero
only over a finite interval $[0,T]$, then a discrete finite time-grid can be
introduced to span this time-interval and also the corresponding finite frequency representation. 
For even $N$ we have:
%
%
\begin{eqnarray}
&t_j&=\frac{jT}{N}, \; j=0,\ldots,N-1 \; ,   \nonumber \\
&\nu_j&=\frac{j}{T}, \; j=0, \ldots,\frac{N}{2}-1 \; ,\nonumber \\
&\nu_{N/2} &= -\frac{N}{2T} \; ,\nonumber \\
&\nu_j&= -\nu_{N-j}, \; j=\frac{N}{2}+1, \dots, N-1.
\label{discretisation1D}
\end{eqnarray}
If the time $t$ and frequencies $\nu$ 
are selected as one of the points of the above discrete grid, 
a discrete version of the integral can be computed.
The continuous Fourier transform is approximated by a discrete Fast-Fourier Transform (FFT):
%
%
\begin{eqnarray}
&&\int_{-\infty}^{+\infty}f(t)e^{-2i\pi\nu_k t} dt = \nonumber \\
&&\int_{0}^{T}f(t)e^{-2i\pi\nu_k t} dt \approx \frac{T}{\sqrt{N}} FFT_k (f) 
\label{FFT1}
\end{eqnarray}
where
%
%
\begin{equation}
FFT_k(f)=\frac{1}{\sqrt{N}}\sum_{j=0}^{N-1}f_j e^{-i2\pi jk/N}
\end{equation}
with $f_j = f(t_j)$. 
The integral in equation \eqref{solution_continue} is then apparently approximated by the sum 
%
%
\begin{equation}
I(t_j)=\frac{1}{2}\frac{T}{\sqrt{N}}FFT_{k=0}(f) - \frac{i}{2\pi}\frac{1}{\sqrt{N}} \sum_{\ell=0}^{N-1} \frac{FFT_{\ell}(f) }{\nu_{\ell}} 
e^{ 2i\pi \ell j/N}.
\label{solution_disc}
\end{equation}
However, the passage from the continuous expression \eqref{solution_continue} to
the FFT equivalent (equation \eqref{solution_disc}) has not been carried out with sufficient rigour, since 
there is a division by $0$ for $\ell=0$ in \eqref{solution_disc}. 
We must adjust the term $\ell = 0$ before doing the inverse transform.
The use of a Cauchy principal value and the presence of a singularity in \eqref{solution_continue}
at $\nu =0$ if $F(\nu=0) \neq 0$ are inconsistent with the use of a Fourier series expansion. 
The equations should thus be modified to take into account this discrepancy. 
However, we can preserve the feature that only two Fourier transforms are required to obtain the integral.

The calculation of the $N$ integrals $I(t_j),\; j=0,\ldots, N-1$ is made as follows. 
The Fourier transform $FFT(f)$ is first calculated and then a new function
 $\tilde{f}(t)$
is defined:
%
%
\begin{equation}
\tilde{f}(t)=f(t)-\frac{1}{\sqrt N} FFT_{k=0}(f).
\label{soustraction}
\end{equation}
Both $f$ and $\tilde{f}$ have the same discrete spectrum, except for the $\ell=0$ component, which for $\tilde{f}$ is equal to zero,
so that the integral of $\tilde{f}(t)$ over the interval $[0,T]$ is equal to zero. 
This is equivalent to using a translation of the ordinate axis. We work temporarily with $\tilde{f}$ instead of $f$. 
We assume that the spectrum $\tilde{F}(\nu)$ (equal to zero for $\nu =0$) has a Taylor expansion near $\nu =0$: 
$$\tilde{F}(\nu)= a \nu +b \nu^2 + \ldots$$ 
We are looking for the finite coefficient 
$a = \lim\limits_{\nu \to 0} \frac{\tilde{F} (\nu) } {\nu}$, so as to apply 
equation \eqref{solution_disc} at $\ell=0$ in a consistent way. 
The divergence produced by the term $\nu_{\ell} \propto \frac{1}{\ell} $ for $\ell=0$ in \eqref{solution_disc}
has disappeared but the value of $a$ is still unknown. 
One can guess it by requiring that $I(t_0=0)=\int_0^0 \tilde{f}(t) dt =0$. 
Equation \eqref{solution_disc} in this case leads to the value 
%
%
\begin{equation}
a =  - \sum_{\ell=1}^{N-1} \mu_{\ell}.FFT_{\ell} (f),
\label{coef_manquant}
\end{equation}
where $\mu.FFT (f)$ denotes a component by component vector multiplication, with
%
%
\begin{equation}
\mu_{\ell}=\left \{ \begin{array}{l}
\frac{i}{2\pi}\frac{T}{\ell} \;\;\;\; \ell=1, \ldots \frac{N}{2}-1  \; ,\\
\frac{i}{2\pi}\frac{T}{\ell-N} \;\;\;\; \ell=\frac{N}{2}, \ldots N-1. \\
\end{array} \right. \\
\label{bons_coefs}
\end{equation}
This value of $a$ replaces the $\ell=0$ component of $(\mu . FFT({f}))$ in equation \eqref{eq_finale_1d} below. 
So far we have ignored the zero frequency component of the function by working with $\tilde{f}$ instead of $f$. 
The final result should then be corrected by a linear correction function, giving 
%
%
\begin{equation}
I(t_j)=FFT_j^{-1}(\mu.FFT(f)) + \frac{j}{N}\times \frac{T}{\sqrt N} FFT_{k=0}(f).
\label{eq_finale_1d}
\end{equation}
The first term on the right hand side represents the integral of $\tilde{f}$ at the discrete time values $t_j$
and the second term restores the correct value of $\int_0^{t_j}f(t) dt$. 
Equations \eqref{eq_finale_1d}, \eqref{bons_coefs} and \eqref{coef_manquant} are the basic equations needed 
to calculate the $N$ integrals $I(t_j)$ using two FFT calls (with a computational cost scaling as $2 N \log N$).

The above equations have close similarities with those of the  
FFT-based integration algorithms proposed in ref. \cite{Yarolavsky2005,Yarolavsky2007} 
in the framework of image resampling. 
The derivation starting from \eqref{solution_continue}
emphasizes the difficulties introduced by the Cauchy principal value term for small frequencies.
Here the principal value integral is approximated by first cancelling $\tilde{F}(\nu = 0)$ and guessing 
its first derivative at $\nu=0$ in a consistent and unambiguous way. 
Our expressions also work well for functions with integrals
over the interval $[0,T]$ which are not necessarily equal to zero.

\subsection{Integration of non-periodic functions \label{int_non_per}}

In the algorithm developed in section \ref{iterative_proc} the smoothness of the wave operator and the wavefunction 
at the time boundary $t=T$ is ensured by the presence of a time-dependent absorbing potential (cf. section \ref{abs_pot}). 
However the intermediate matrix elements of $H_{eff}^{(n)}(t)$ and $\tilde{H}^{(n)}_{diag,kk}(t)$ to be integrated are not necessarily T-periodic. 
FFT-based differentiation and integration algorithms 
are very accurate 
but suffer from boundary effects (Gibb's phenomenon). 
The present algorithm will give 
accurate results only with periodic continuous functions, ideally
those with time derivatives 
which are themselves also continuous functions. 
There are different ways of adapting a Fourier-based algorithm to deal with non-periodic signals. 
For example, extension methods have been developed to solve partial differential equations on a complex domain \cite{elghaoui1996,averbuch1997}, 
to represent surfaces of complex bodies \cite{bruno2007}
and to treat imaging problems \cite{Yarolavsky2005}. 

Here we use the alternative procedure of Hermite interpolants
to ensure the continuity of the function $f(t)$ and 
of the first derivatives \cite{boyd2002}.  
Let $f(t)$ be a complex continuous $L^2$ function defined on a large time interval $[0,T_0]$, and represented 
by a sample of $N_0$ equally distributed values $f(t_j),\;\; t_j=j T_0/N_0,\; j=0,\ldots, N_0-1$ on this interval. 
The procedure described in \eqref{soustraction} has been applied to $f(t)$ so that $FFT_{k=0}(f)=0$.
As the function $f(t)$ is assumed to be continuous with continuous first and second derivatives inside
the interval $[0,T_0]$, the principal problem is to ensure 
continuity at the boundaries. 
Let $f(0), f'(0), f''(0)$ and $f(T_0), f'(T_0), f''(T_0)$ be the values of the function and of the first two
derivatives at the two boundaries. 
It will be impossible to obtain accurate results if these values do not match. 
One can extend 
the time interval by adding a narrow extra interval $[T_0,T]$
spanned by an extension of the grid $t_j=j T_0/N_0,\; j=N_0, \ldots, N-1$ and then constructing on this new interval a 
$6^{th}$ order polynomial:
%
%
\begin{equation}
f(t)=\sum_{j=0}^6 a_j (t-T_0)^j.
\label{polynome}
\end{equation}
Seven conditions
are imposed on the polynomial. The first three are related to the continuity of $f(t)$ at the point 
$t_j=T_0$ and give the first three coefficients $a_j$:

%
%
\begin{equation}
\begin{array}{l}
a_o=f(T_0), \\
a_1=f'(T_0), \\
a_2=f''(T_0)/2.
\end{array} 
\label{coef_pol}
\end{equation} 
The three values on the right hand sides of \eqref{coef_pol} are known exactly if the function $f(t)$ has an analytical expression.
Alternatively, the derivatives can be approximated by a FFT algorithm if $f(t)$ is known at a sample of $N_0$ values. 
The last four coefficients 
are related to the continuity between $t=T$ and $t=0$  
of $f(t)$ and to the condition $\int_0^{T}f(t) dt=0$: 
%
%
\begin{equation}
\begin{array}{l}
f(T)=f(0), \\
f'(T)=f'(0) ,\\
f''(T)=f''(0), \\
\int_{T_0}^{T} f(t) dt =0.
\end{array} 
\label{continuite}
\end{equation} 
With $\Delta=T-T_0$, the last four unknowns
are found by solving the small linear system:
%
%
\bea
 \left(
\begin{array}{cccc}
\Delta^3 & \Delta^4 &\Delta^5 & \Delta^6 \\
3\Delta^2 & 4\Delta^3 & 5\Delta^4 & 6 \Delta^5 \\ 
6\Delta & 12\Delta^2 & 20 \Delta^3 & 30 \Delta^4 \\
\frac{\Delta^4}{4} & \frac{\Delta^5}{5} & \frac{\Delta^6}{6} & \frac{\Delta^7}{7} \\
\end{array}
\right)  
\left(
\begin{array}{c}
a_3 \\
a_4 \\
a_5 \\
a_6 \\
\end{array}
\right) \nonumber \\
= 
\left(
\begin{array}{c}
f(0) -a_0-a_1\Delta-a_2\Delta^2 \\
f'(0) -a_1-2 a_2\Delta \\
f''(0) -2a_2 \\
-a_0\Delta-a_1\frac{\Delta^2}{2}- a_2\frac{\Delta^3}{3} \\ 
\end{array}
\right)
\label{systeme}
\eea
Note that the last (integral) condition \eqref{continuite} is not absolutely necessary and can be relaxed 
by first defining the above polynomial and then applying the zero-integral condition on the whole interval using equation \eqref{soustraction}. 

In summary, the FFT-based integration algorithm is implemented using \eqref{coef_manquant}-\eqref{eq_finale_1d}, 
completed by a preliminary treatment on
the additional interval $[T_0,T]$ using \eqref{polynome}-\eqref{systeme}. 

\end{document}